\documentclass[usenatbib,final]{mn2e}
\usepackage{graphicx,tabularx,amsmath,amssymb,upgreek,wasysym,mathtools,multirow,url,rotating,pdflscape,color,soul}
\numberwithin{equation}{section}

\graphicspath{{figures/}}

\newcommand{\change}[1]{#1}

\title[The structure of young clusters]{$\mathcal{Q}^{+}$: Characterising the structure of young star clusters}
\author[S. E. Jaffa, A.d P. Whitworth \& O. Lomax]{S. E. Jaffa\thanks{E-mail: Sarah.Jaffa@astro.cf.ac.uk}, A. P. Whitworth, O. Lomax \\  
School of Physics and Astronomy, Cardiff University, Cardiff CF24 3AA, Wales, UK}

\begin{document}

\newpage
\pagerange{\pageref{firstpage}--\pageref{lastpage}} \pubyear{2013}
\maketitle
\label{firstpage}

\begin{abstract} 
\change{Many young star clusters appear to be fractal, i.e. they appear to be concentrated in a nested hierarchy of clusters within clusters. We present a new algorithm for statistically analysing the distribution of stars to quantify the level of sub-structure. 
We suggest that, even at the simplest level, the internal structure of a fractal cluster requires the specification of three parameters. (i) The 3D fractal dimension, $\mathcal{D}$, measures the extent to which the clusters on one level of the nested hierarchy fill the volume of their parent cluster. (ii) The number of levels, $\mathcal{L}$, reflects the finite ratio between the linear size of the large root-cluster at the top of the hierarchy, and the smallest leaf-clusters at the bottom of the hierarchy. (iii) The volume-density scaling exponent, $\mathcal{C}=-\textrm{d}\ln[\delta n]/\textrm{d}\ln[L]$ measures the factor by which the excess density, $\delta n$, in a structure of scale $L$, exceeds that of the background formed by larger structures; it is similar, but not exactly equivalent, to the exponent in Larson's scaling relation between density and size for molecular clouds.
We describe an algorithm which can be used to constrain the values of $({\cal D},{\cal L},{\cal C})$ and apply this method to artificial and observed clusters. We show that this algorithm is able to reliably describe the three dimensional structure of an artificial star cluster from the two dimensional projection, and quantify the varied structures observed in real and simulated clusters.}
\end{abstract}


\section{Introduction}%

The ${\cal Q}$-parameter (\citealt{CartWhit2004}; hereafter CW04) \change{uses the Complete Graph and Minimum Spanning Tree to quantify the structure of a star cluster, and can separate fractal, sub-structured distributions from those with a radial density gradient. It} has frequently been used to characterise the three-dimensional structure of observed and simulated star clusters.\footnote{We use the term `cluster' here generically, to embrace any collection of stars formed in close proximity, thus also including associations and groups.} For example, it has been applied to quantifying the internal structure of nearby newly-formed clusters like Ophiuchus, Taurus, IC348, Chamaeleon, IC2391, Serpens and Auriga-California (CW04, \citealt{SchmKles2006, Schmetal2008, Broeetal2014}), more distant larger newly-formed clusters like W40, RCW 38, AFGL 490, LDN 1188, Cyg OB2, W5-east, NGC7538, S235, S252, S254-S258, NGC6334, Carina-west \citep{Kuhnetal2010, Winsetal2011, Masietal2012, Wrigetal2014, Chavetal2014, Huntetal2014, Kumaetal2014}, open clusters \citep{SancAlfa2009, Fernetal2012, Delgetal2013, Gregetal2015}, globular clusters \citep{Beccetal2012}, and clusters in the Magellanic Clouds \citep{Schmetal2009, Valletal2010, Gouletal2012, Gouletal2014a, Gouletal2014b}.

${\cal Q}$ has also been used to look for signatures of sequential star formation and mass segregation in observed star clusters \citep[e.g.][]{KumaSchm2007, Caballer2008, Allietal2009a, Kuepetal2011, Camaetal2011, Gagnetal2015}, the distribution of cores in the Galaxy \citep{PlanckCo2011}, the evolution of stellar distributions in the Magellanic Clouds and other external galaxies \citep{Gieletal2008, Bastetal2009, Gouletal2010, Bastetal2011, Hascetal2012, Gouletal2015}, and even the distribution of field objects near radio galaxies \citep{KeshVerk2015}.

Finally, ${\cal Q}$ has been used to analyse the output from simulations, in particular, the underlying structure of star clusters formed in simulations \citep{Kirketal2014,Balfetal2015}, their dynamical evolution \citep{MoecBate2010, Mascetal2010, Allietal2010, Smitetal2011, Girietal2012b, ParkMeye2012, Parketal2014, ParkerRJ2014}, their response to feedback from massive stars \citep{Daleetal2012, Daleetal2013, ParkDale2013, Parketal2015, ParkDale2015}, and their degree of mass segregation \citep{Allietal2009b, ParkGood2015}.

${\cal Q}$ is evaluated by first constructing the Complete Graph of a two-dimensional set of points (e.g. the projected positions of stars in a cluster) and computing the mean length, $\bar{s}$, of all the edges on the Complete Graph (i.e. all the straight lines connecting each point to all the other points). Next, one constructs the Minimum Spanning Tree (MST) of the points and computes the normalised mean length, $\bar{m}$, of the edges on the MST. Finally, one computes ${\cal Q}=\bar{m}/\bar{s}$. Values of ${\cal Q}<0.8$ can be translated into a notional fractal dimension, ${\cal D}$, for star clusters with substructure, and values of ${\cal Q}>0.8$ can be translated into a notional radial density exponent, $\alpha = - \change{\textrm{d}}\ln[n]/\change{\textrm{d}}\ln[r]$ for spherically symmetric star clusters (here $n$ is the mean volume-density of stars at distance $r$ from the centre of the cluster). ${\cal Q}$ can also be evaluated for continuum images of clouds, for example long-wavelength Herschel maps of molecular clouds \citep{Lomaetal2011,ParkDale2015}. To do this, the continuum image must be converted into an ensemble of points.

However, even if star clusters are fractal, their fractal dimension, ${\cal D}$, does not fully capture the statistics of their internal structure. One needs to specify the range of spatial scales \change{(in the context of turbulence this is sometimes called the 'inertial range')} over which the cluster is fractal, i.e. the ratio $2^{\cal L}$ between the overall linear size of the cluster and the smallest sub-sub-...-sub-cluster, not least because a cluster with a finite number of stars can only populate a finite range of scales. One also needs to specify the extent to which the stars are concentrated in the smaller scales of the hierarchy, i.e. a volume-density scaling exponent, ${\cal C}=-\change{\textrm{d}}\ln[\delta n_\ell]/\change{\textrm{d}}\ln[L_\ell]$, where $\delta n_\ell$ is the additional volume-density in, and $L_\ell$ the mean linear size of, the clusters on level $\ell$ of the hierarchy. If ${\cal C}$ is low, the smaller clusters constitute a very small density excess relative to the background defined by the larger clusters, whereas if ${\cal C}$ is high, most of the stars are in the smaller clusters and their background is relatively sparse. \change{More detailed definitions of these three parameters are given in Section \ref{SEC:MAKE}. We note that the $\mathcal{Q}$-parameter defined in \citet{CartWhit2004} is restricted by only considering ${\cal D}$ explicitly, and implicitly adopting the defaults ${\cal L} = \log_{2}(\mathcal{N}_\star^{1/{\cal D}})$ (where $\mathcal{N}_\star$ is the total number of stars) and ${\cal C}=\infty$.}

It follows that more sophisticated measures than ${\cal Q}$ are required, and we attempt to develop such measures here. The paper is organised as follows. In Section \ref{SEC:MAKE} we present a procedure for generating synthetic fractal star clusters characterised by ${\cal D}$, ${\cal L}$ and ${\cal C}$ (a more sophisticated procedure will be presented in Whitworth \& Jaffa, in prep.). In Section \ref{SEC:COMP} we illustrate (projected) clusters constructed using this procedure, and how their properties are influenced by varying ${\cal D}$, or ${\cal L}$, or ${\cal C}$. In Section \ref{SEC:METHOD} we explain how complete graphs and Minimum Spanning Trees are constructed, and define the discriminating measures that can be derived from them. We demonstrate how these measures can be combined to express the maximum variation with ${\cal D}$, ${\cal L}$ and ${\cal C}$, and we explain how estimates of ${\cal D}$, ${\cal L}$ and ${\cal C}$ can be inferred from the projected image of a real star cluster. In Section \ref{SEC:APPLY} we apply the algorithm to both synthetic data and observed star clusters, to evaluate its reliability and compare the results with those obtained previously using ${\cal Q}$. In Section \ref{SEC:CONC} we summarise our main conclusions.

For mathematical convenience, we define a 3-component statistical state vector for a fractal star cluster,\newline ${\bf Y}\equiv (\mathcal{D}, \mathcal{L}, \mathcal{C})$.

\begin{figure*}
\centering
\includegraphics[width=\linewidth]{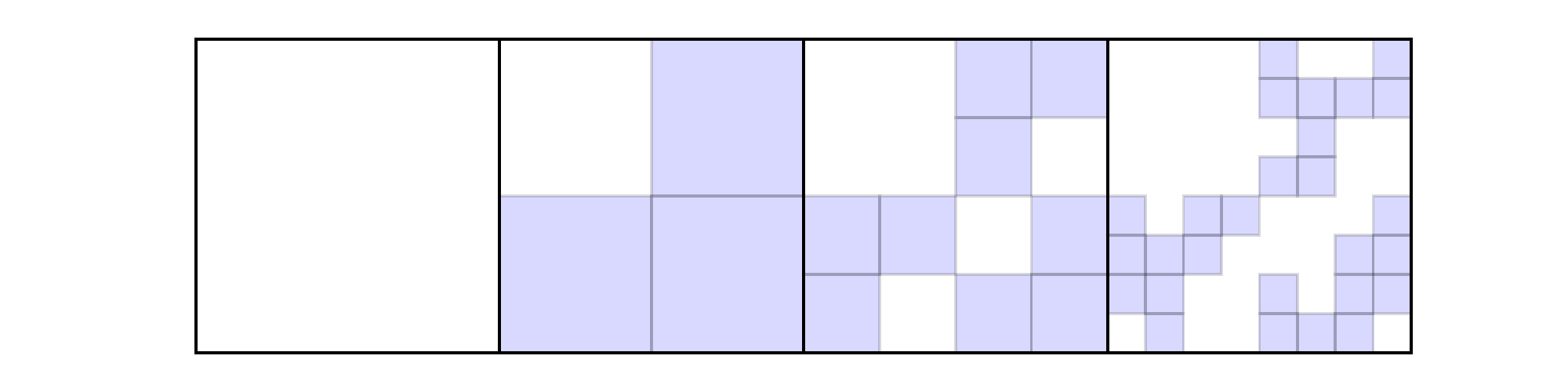}
\caption{A square two-dimensional field partitioned with (two dimensional) fractal dimension $D_{_2}=1.585$. Note that (a) the root-square on level $\ell=0$ is shaded, because it is fertile; (b) three of the squares on level $\ell=1$ are shaded, because they are fertile; (c) nine of the squares on level $\ell=2$ are shaded, because they are fertile; (d) twenty seven of the squares on level $\ell=3$ are shaded, because they are fertile. Note that in this case we are using the two-dimensional fractal dimension, defined similarly to equation \ref{P_fertile} but with a 2 in the power instead of 3, i.e. ${\cal P}_{_{\rm FERTILE}}=2^{({\cal D}-2)}$}
\label{FIG:2DFIELD}
\end{figure*}

\section{Constructing synthetic three-dimensional fractal star clusters}\label{SEC:MAKE}%

The algorithm to construct a synthetic three-dimensional fractal star cluster uses recursive octal partitioning, following \cite{2004G+W}. It starts with a root-cube of side $L_{_0}=2$, centred on the origin, i.e. $-1<x,y,z<+1$; the root-cube constitutes level $\ell=0$, and it is {\it de facto} `fertile' (see below).

\subsection{The fractal dimension, ${\cal D}$}\label{SSEC:D3}

The root-cube is divided into 8 equal cubes, each of side $L_{_1}=1$, and a random subset of these cubes is tagged as being fertile. The probability of a given cube being fertile is given by
\begin{equation}
{\cal P}_{_{\rm FERTILE}}=2^{({\cal D}-3)}\,,
\label{P_fertile}
\end{equation}
so reducing ${\cal D}$ decreases the probability of a cube being fertile. Cubes that are not fertile are sterile, and play no further part. The fertile cubes constitute level $\ell=1$.

Each fertile cube is then divided into 8 equal sub-cubes, each of side $L_{_2}=0.5$, and a random subset of these sub-cubes is tagged as being fertile. The probability of being fertile is again ${\cal P}_{_{\rm FERTILE}}$, and any sub-cubes that are not fertile are sterile, and play no further part. The fertile sub-cubes constitute level $\ell=2$.

This procedure is repeated recursively, so that at each level, $\ell$, each fertile parent-cube on level $\ell$ is divided into 8 child-cubes on level $\ell+1$, and these child-cubes have a probability ${\cal P}_{_{\rm FERTILE}}$ of being fertile and therefore spawning grandchild-cubes on the next level, $\ell+2$. See Figure \ref{FIG:2DFIELD} for a two dimensional demonstration of this procedure, with 3 out of 4 sub-squares being fertile at each division.\footnote{We stress that this demonstration is intrinsically two-dimensional solely because it is easier to illustrate on paper. In the preceding sections of the paper, and in what now follows, we discuss exclusively three-dimensional clusters, although we are concerned with how one interprets their appearance when they are seen from only one direction, projected on the sky.}

The best behaved results are obtained when ${\cal F}=2^{{\cal D}}$ is an integer, since each division then simply involves choosing randomly -- from 8 child-cubes -- the ${\cal F}$ child-cubes that are fertile. We therefore limit the artificial clusters generated to integer values of ${\cal F}$.

\subsection{The number of levels, ${\cal L}$}\label{SSEC:R}
The root-cube is labelled as level 0 and at each splitting the fertile parent cubes are split into 8 children. The recursive division is terminated at level $\mathcal{L}$, as soon as we have created a level of sub-sub-...-sub-cubes which are smaller than the root-cube by a factor $\mathcal{R}$, so
\begin{equation}
\mathcal{L} = \log_{2}(\mathcal{R})
\end{equation}
The root-cube has side 2, and the cubes on level $\ell$ have side $2^{1-\ell}$, so the range of sizes is $2^{\cal L}={\cal R}$.
The sub-sub-...-sub-cubes on the final level ${\cal L}$ are termed the leaf-cubes. The two dimensional case shown in figure \ref{FIG:2DFIELD} has 3 levels.

\subsection{The volume-density scaling exponent, ${\cal C}$}\label{SSEC:C3}

We define the volume-density initially assigned to the root-cube as $n_{_0}$. The additional volume-density assigned to the fertile cubes on level $\ell=1$ is then $\delta n_{_1}=n_{_0}2^{\,\mathcal{C}}$. The additional volume-density assigned to the fertile sub-cubes on level $\ell=2$ is $\delta n_{_2}=\delta n_{_1}2^{\,\cal C}=n_{_0}2^{2\mathcal{C}}$. Sub-sub-...-sub-cubes on level $\ell$ are assigned additional volume-density $\delta n_\ell=n_{_0}2^{\ell{\cal C}}$.

The volume of space occupied by {\it all} the fertile cubes on level $\ell$ is $V_g=8{\cal P}_{_{\rm FERTILE}}^\ell$, and hence the total number of stars in the root-cube (including stars assigned to the smaller cubes that are its descendants in the hierarchy) is

\begin{eqnarray}\nonumber
{\cal N}_{_{\rm ROOT\;CUBE}}&=&V_{_0}n_{_0}\,+\,\sum\limits_{i=1}^{\mathcal{L}}V_{_i}\delta n_{_i} \\\nonumber
&=&8n_{_0}\,\sum\limits_{i=0}^{\mathcal{L}}\mathcal{B}^{i} \\
&=&\frac{8n_{_0}\left({\cal B}^{({\cal L}+1)}-1\right)}{\left({\cal B}-1\right)}\,,\\
{\cal B}&=&2^{({\cal C}+{\cal D}-3)}\,.
\end{eqnarray}

The additional number of stars in a single fertile leaf-cube on the last level is $n_{_0}2^{3+{\cal L}({\cal C}-3)}$, and this must be unity, so
\begin{eqnarray}
n_{_0}&=&2^{({\cal L}(3-{\cal C})-3)}\,,\\
{\cal N}_{_{\rm ROOT\;CUBE}}&=&\frac{2^{({\cal L}(3-{\cal C}))}\left({\cal B}^{({\cal L}+1)}-1\right)}{\left({\cal B}-1\right)}\,.
\end{eqnarray}

Each fertile cube on level $\ell$ is therefore allocated $\delta {\cal N}_\ell=2^{(\ell-\mathcal{L})({\cal C}-3)}$ stars, which are positioned randomly within the cube. Non-integer numbers of stars are accommodated with a cumulative remainder.

Finally, the root-cube is pruned to a sphere with radius $R=1$, and rotated through random Euler angles. The total number of stars in the cluster is therefore
\begin{equation}
{\cal N}_\star\simeq\frac{\pi}{6}\;\frac{2^{({\cal L}(3-{\cal C}))}\left({\cal B}^{({\cal L}+1)}-1\right)}{\left({\cal B}-1\right)}\,.
\label{EQN:NSTAR}
\end{equation}

The number of stars in a cluster increases with increasing $\mathcal{D}$, increasing $\mathcal{L}$, and decreasing $\mathcal{C}$.

\section{The qualitative effects of changing ${\cal D}$, ${\cal L}$ or ${\cal C}$}\label{SEC:COMP}%

\begin{figure*}
\centering
\includegraphics[width=2\columnwidth]{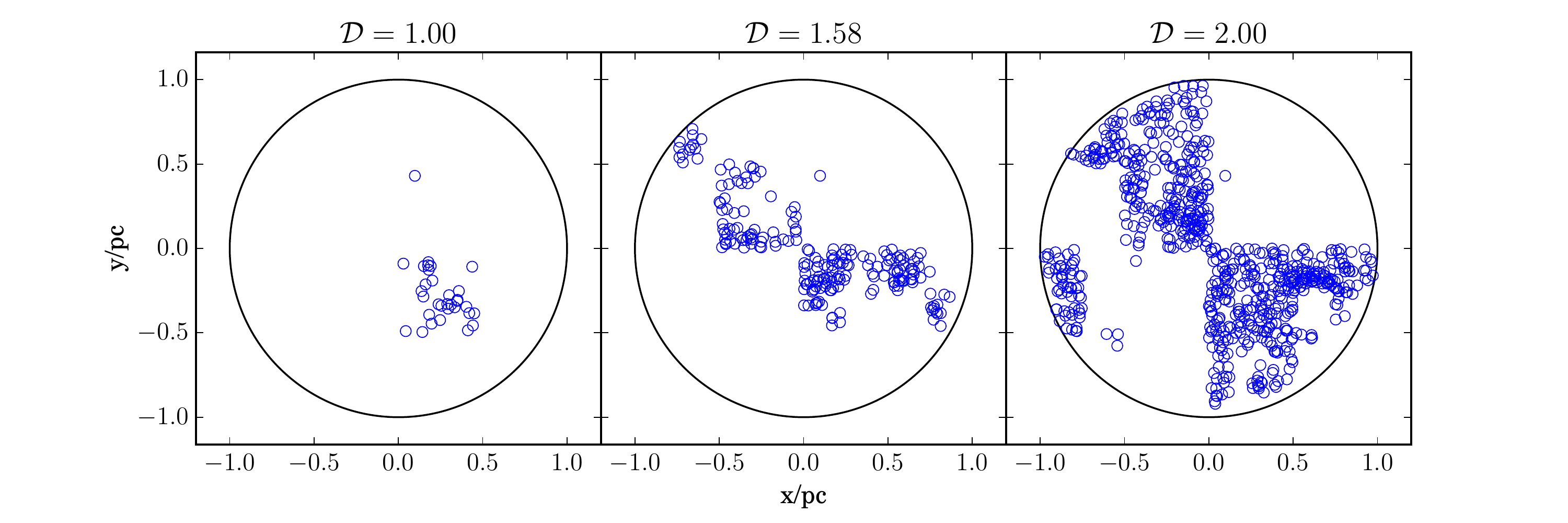}
\caption{2D projections of 3D fractal star clusters, all with the same ${\cal L}=5$ and ${\cal C}=3$, but different ${\cal D}$. (a) ${\cal D}=1.00$; (b) ${\cal D}=1.58$ (the fiducial case); (c) ${\cal D}=2.00$. Increasing ${\cal D}$ reduces the amount of empty space in the cluster and increases the number of stars.}
\label{FIG:VaryingD3}
\end{figure*}

\begin{figure*}
\centering
\includegraphics[width=2\columnwidth]{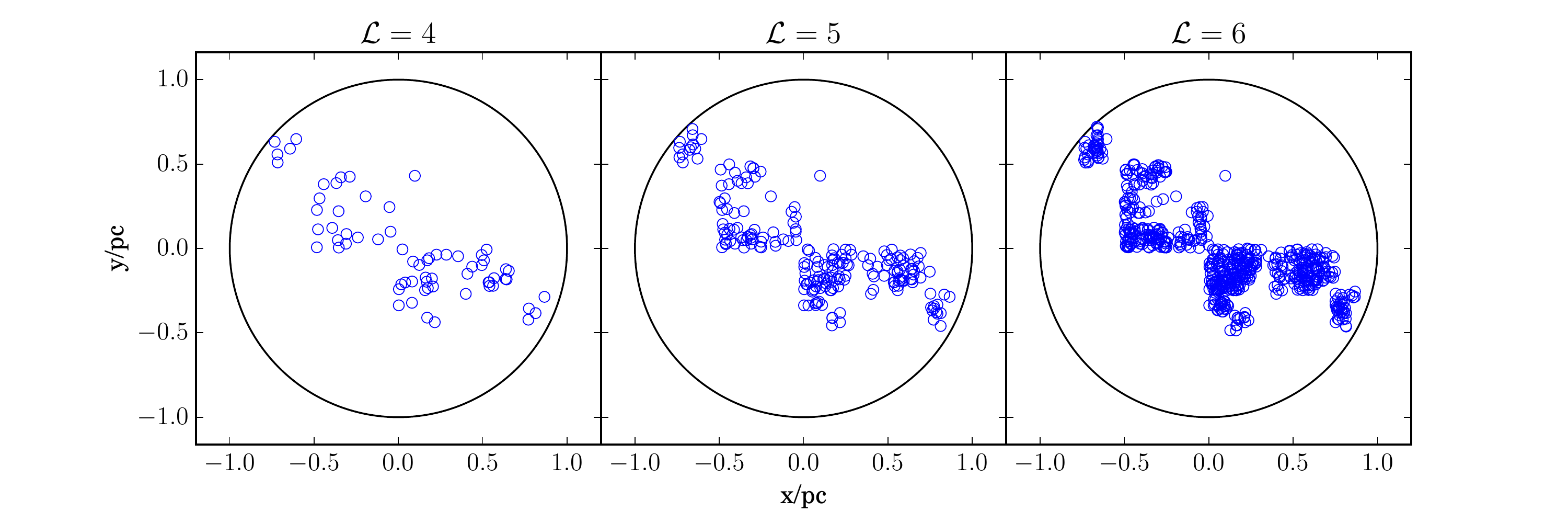}
\caption{2D projections of 3D fractal star clusters, all with the same ${\cal D}=1.58$ and ${\cal C}=3$, but different ${\cal L}$. (a) ${\cal L}=4$; (b) ${\cal L}=5$ (the fiducial case); (c) ${\cal L}=6$. Increasing ${\cal L}$ decreases the size of the smallest separations compared to the overall size of the cluster and increases the number of stars.}
\label{FIG:VaryingR}
\end{figure*}

\begin{figure*}
\centering
\includegraphics[width=2\columnwidth]{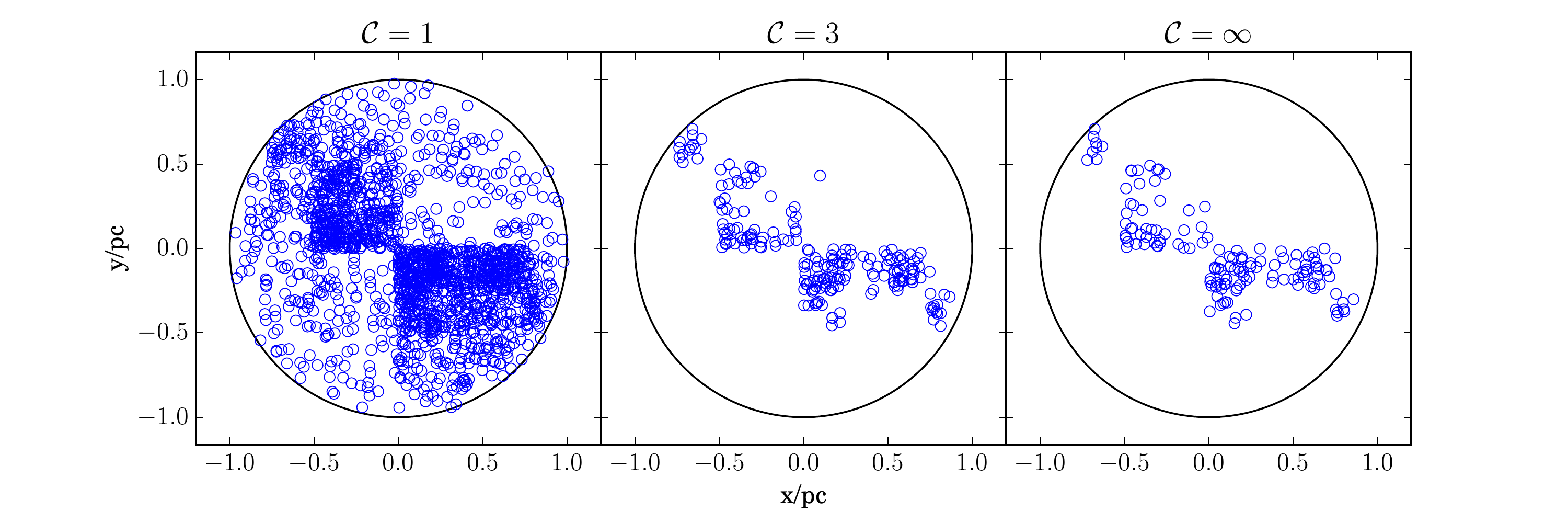}
\caption{2D projections of 3D fractal star clusters, all with the same ${\cal D}=1.58$ and ${\cal L}=5$, but different ${\cal C}$. (a) ${\cal C}=1$; (b) ${\cal C}=3$; (c) ${\cal C}=\infty$. Increasing $\cal C$ concentrates the star more on the later generations (smaller structures) of the fractal and decreases the number of stars.}
\label{FIG:VaryingC3}
\end{figure*}

In this section we illustrate three-dimensional clusters generated using the algorithm described in Section \ref{SEC:MAKE} and projected onto the plane of the sky, in order to demonstrate the effect of varying the underlying parameters, ${\cal D}$, ${\cal L}$ and ${\cal C}$. For reference we define a fiducial cluster with ${\cal D}=1.58$ , ${\cal L}=5$, and ${\cal C}=3$. Figures \ref{FIG:VaryingD3}, \ref{FIG:VaryingR} and \ref{FIG:VaryingC3} show representative randomly generated clusters that have not been rotated. Viewing along the Cartesian axes we can more clearly identify the structural influence of each parameter.

\subsection{The effect of changing the fractal dimension, ${\cal D}$.}

Fig. \ref{FIG:VaryingD3} shows clusters with three different values of ${\cal D}$, but the same ${\cal L}=5$ and ${\cal C}=3$. The left hand image shows a cluster with ${\cal D}=1.00$; in this case the fractal dimension is low, and in the partitioning of space each parent cube has only 2 fertile child-cubes (plus 6 sterile ones), so the cluster is very sparse. The middle image shows a cluster with ${\cal D}=1.58$; this is the fiducial cluster with a middling fractal dimension, and each parent-cube has 3 fertile child-cubes (plus 5 sterile ones), so the cluster is more uniformly populated. The right hand image shows a cluster with ${\cal D}=2.00$; this is a higher fractal dimension, and each parent-cube has 4 fertile child-cubes and 4 sterile child-cubes, so the cluster is populated more uniformly. Thus the effect of increasing $\mathcal{D}$ is to increase the volume-filling factor on every level.

\subsection{The effect of changing the number of levels, ${\cal L}$}

Fig. \ref{FIG:VaryingR} shows clusters with three different values of ${\cal L}$, but the same ${\cal D}=1.58$ and ${\cal C}=3$. The left hand image shows a cluster with ${\cal L}=3$. The middle image again shows the fiducial cluster with ${\cal L}=5$. The right hand image shows a cluster with ${\cal L}=7$. The broad structures seen in the higher-${\cal L}$ cases are visible, but less clearly defined, in the lower-${\cal L}$ case.

\subsection{The effect of changing the volume-density scaling exponent, ${\cal C}$}

Fig. \ref{FIG:VaryingC3} shows clusters with three different values of ${\cal C}$, but the same ${\cal D}=1.58$ and ${\cal L}=5$. The left hand image shows a cluster with ${\cal C}=1$; this is a small scaling exponent, which means that the excess volume-density in a child-cluster is not much greater than that of its parent-cluster (i.e. the substructure is not very well defined). The middle image again shows the fiducial cluster with ${\cal C}=3$, with child-clusters slightly denser than their parent-clusters. The right hand image shows a cluster with ${\cal C}=\infty$, where all the stars are located in the leaf-cubes on the final level. When ${\cal C}=1$ the many stars on the first level swamp any substructure on the lower levels. We therefore concentrate on clusters with ${\cal C}>1$.

\section{The $\mathcal{Q}^+$ algorithm}\label{SEC:METHOD}%

Given a 2D image of a star cluster containing ${\cal N}_\star$ stars, we seek to constrain the parameters, ${\bf Y}$, describing its intrinsic 3D structure. Implicitly, we assume that the intrinsic 3D structure conforms to the fractal model described in Section \ref{SEC:MAKE}.

We define a set of discriminating measures that distinguish the 3 parameters of an artificial fractal model based on the Minimum Spanning Tree and the Complete Graph (see section \ref{SUBSEC:MEAS}). $\mathcal{D}$ and $\mathcal{L}$ influence many of these measures so we combine them using Principle Component Analysis (see section \ref{SUBSEC:PCA}). However, $\mathcal{C}$ has a more subtle effect on the structure and does not strongly influence many of the measures. We therefore treat this parameter separately once $\mathcal{D}$ and $\mathcal{L}$ have been estimated (see section \ref{SUBSEC:BAYES}).

We consider star clusters with the following properties, \\

\begin{tabular} {rllllll}
${\cal D}=\;\;$ & 1.00, & 1.58, & 2.00, & 2.32; & \textcolor{white}{0.00}&  \\
${\cal L}=\;\;$ & 3, & 4, & 5, & 6, & 7, & 8; \\
${\cal C}=\;\;$ & 1, & 2, & 3, & $\infty$. &&\ \\
\end{tabular}\\

\noindent There are 96 combinations in total. For each tabulated ${\bf Y}=({\cal D},{\cal L},{\cal C})$, we calculate the expected number of stars (Equation \ref{EQN:NSTAR}). We exclude clusters whose ${\cal N}_{\star}$ would give too few stars for statistical significance or too many for computational efficiency, leaving 65 \textbf{Y} states with  $20 \leqslant {\cal N}_{\star} \leqslant 10000$. For these parameters, we generate 100 independent star clusters and compute several possible measures that could distinguish the cluster structure.

\subsection{Measures derived from Complete Graphs and Minimum Spanning Trees}\label{SUBSEC:MEAS}%

We first construct the Complete Graph, i.e. the collection of ${\cal N}_{\star}({\cal N}_{\star}-1)/2$ edges (straight lines) connecting each star with every other star. The length of the edge joining stars $i$ and $j$ on the Complete Graph is termed $s_{ij}$.

Next we construct the Minimum Spanning Tree (MST), i.e. the collection of ${\cal N}_{\star}-1$ edges that connects every star directly to at least one other star and thereby indirectly to all other stars \change{with no closed loops}, and has the minimum total length. The length of the $k^{\rm th}$ shortest edge on the MST is termed ${m_k}$.

We assume that the cluster is spherical, and therefore its projection is circular, with radius $R$. We do not introduce the notion of a convex hull \citep[c.f.][]{SchmKles2006}, since a cluster generated on the assumption of spherical symmetry, but with a low fractal dimension, ${\cal D}$, and/or a high volume-density scaling exponent, ${\cal C}$, can have an extremely elongated convex hull (see Figs. \ref{FIG:VaryingD3}, \ref{FIG:VaryingR} and \ref{FIG:VaryingC3}); it still belongs to a family of clusters built upon the assumption of spherical symmetry.

The 7 statistical measures that are most useful in classifying the structure are the following:

\begin{enumerate}
\item the logarithm of the number of stars,
\begin{equation}
{\log(\mathcal{N}_{\star})};
\end{equation}

\item the logarithm of the range of edges on the Complete Graph, $\log(\mathcal{R})$, which is given by
\begin{equation}
{\log(\mathcal{R})}=\frac{s_{_{\rm MAX}}}{s_{_5}},
\end{equation}
where $s_{_{\rm MAX}}$ is the largest edge on the Complete Graph, and $s_{_5}$ is the fifth smallest;\footnote{\change{Whereas the largest edge on the Complete Graph, $s_{_{\rm MAX}}$, is relatively robust -- in the sense that, if ($\mathcal{D, L, G}$) are held constant, it varies very little from one realisation to another -- the smallest edge, $s_{_{\rm MIN}}$, is not. $s_{_{\rm MIN}}$ has a large variance because it is usually determined by one chance alignment, and therefore can be arbitrarily small. We mitigate this problem by using the fifth smallest edge, $s_{_5}$. In the same spirit, \citet{2009L} used the 5th brightest cluster in a galaxy as representative of the absolute magnitude, and this practice is often used in extragalactic statistics.}}

\item the normalised mean edge length on the Minimum Spanning Tree, ${\bar m}$, which is given by
\begin{equation}
{\bar m}=\frac{(\mathcal{N}_\star-1)}{(\pi{\cal N}_\star)^{1/2}R}\,\sum\limits_{k=1}^{k={\cal N}_\star-1}\,\left\{m_k\right\};
\end{equation}

\item the normalised mean edge length on the Complete Graph, ${\bar s}$, which is given by
\begin{equation}
\bar{s}=\frac{2}{{\cal N}_\star({\cal N}_\star-1)R}\,\sum\limits_{i=1}^{i={\cal N}_\star-1}\,\sum\limits_{j=i+1}^{j={\cal N}_\star}\,\left\{s_{ij}\right\}\,;
\end{equation}

\item the mean of the edge lengths on the Minimum Spanning Tree, ${\mu_{m}}$, which is given by
\begin{equation}
{\mu_{m}}=\frac{1}{\mathcal{N}_{\star}-1}\,\sum\limits_{k=1}^{k={\cal N}_\star-1}\,\left\{m_k\right\}\,;
\end{equation}

\item the standard deviation of the edge lengths on the Minimum Spanning Tree, ${\sigma_{m}}$, which is given by
\begin{equation}
{\sigma_{m}^2}=\frac{1}{\mathcal{N}_{\star}-1}\,\sum\limits_{k=1}^{k={\cal N}_\star-1}\,\left\{(m_k-{\bar m})^2\right\}\,;
\end{equation}

\item the area above the cumulative distribution of MST edges, $A$, normalised by the number of stars and the size of the cluster.   This is given by
\begin{equation}
A=1-\frac{m_{0} + m_{_{\mathcal{N}_{\star}-1}} + 2\sum\limits_{k=1}^{\mathcal{N}_{\star}-2}m_{_{k}}} {2s_{max}(\mathcal{N}_{\star}-1)}\,,
\end{equation}
\change{and reflects the proportion of very short  edges on the MST (see section \ref{SUBSEC:BAYES}).}
\end{enumerate}

\begin{figure*}
\centering
\includegraphics[width=1.8\columnwidth]{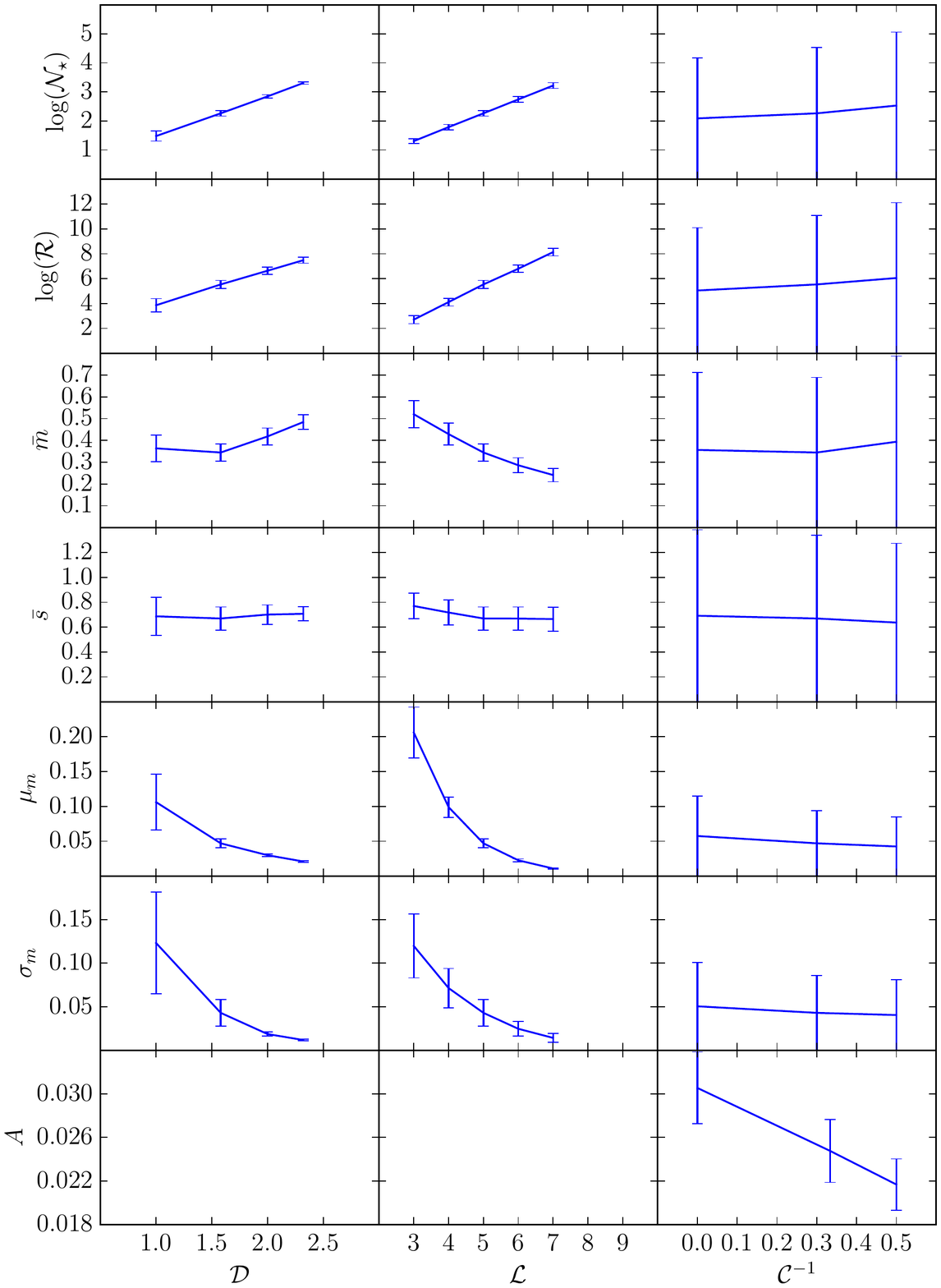}
\caption{The variation of the statistical measures, ($\log(\mathcal{N}_{\star})$ (first/top row), $\log(\mathcal{R})$ (second row), ${\bar m}$ (third row), $\bar{s}$ (fourth row), $\mu_{m}$ (fifth row), $\sigma_{m}$ (sixth row), and $A$ (seventh/bottom row)) with the parameters defining the statistical state, (${\cal D}$ (left hand column), ${\cal L}$ (middle column), and \change{${\cal C}^{-1}$} (right hand column)). The line gives the mean and the error bar gives the standard deviation. Whichever statistical-state parameter is being varied, the other two are held constant at their fiducial values, viz. ${\cal D}=1.58$ , ${\cal L}=5$, and ${\cal C}=3$.}
\label{FIG:MEAS}
\end{figure*}

Fig. \ref{FIG:MEAS} shows how the means and standard deviations of the 7 statistical measures ($\log(\mathcal{N}_{\star}), \log(\mathcal{R}), {\bar m}, \bar{s}, \mu_{m},\sigma_{m}, A$), vary when each of the parameters defining the statistical state, ${\bf Y}$, is varied, with the other two held constant at their fiducial values, ${\cal D}=1.58$, ${\cal L}=5$, and ${\cal C}=3$. The frames in the left hand column show what happens when ${\cal D}$ is varied. The frames in the middle column show what happens when ${\cal L}$ is varied. And the frames in the right hand column show what happens when ${\cal C}$ is varied. The plotted points are the means, and the error bars represent the standard deviations.

\subsection{Estimating $\mathcal{D}$ and $\mathcal{L}$}\label{SUBSEC:PCA}%

Principle component analysis is a mathematical technique first introduced by \citet{pearson1901} for reducing the number of dimensions in data sets with many variables. Using the eigenvectors of the covariance matrix, linear combinations of the dimensions are found which give an orthogonal set of axes which can be used to emphasise the variance in the data, and therefore better separate structures \citep{NUM-RECIP}.

Using many initial parameters, this method was used to find the six most useful measures, which were then combined to give just two orthogonal \textquotedblleft Principal Components" (henceforth referred to as PC 1 and PC 2):

\begin{equation}
\begin{pmatrix}
PC1\\
PC2
\end{pmatrix}=
\begin{pmatrix}
-0.354 & -0.832\\
-0.934 & 0.300\\
0.026 & -0.417\\
0.022 & -0.193\\
0.031 & -0.019\\
0.020 & 0.082
\end{pmatrix}^T
\times
[\begin{pmatrix}
\mathbf{Z}
\end{pmatrix}
-
\begin{pmatrix}
\mathbf{\bar{Z}}
\end{pmatrix}]^T
\end{equation}

\noindent where ${\bf Z}=(\log(\mathcal{N}_{\star}), \log(\mathcal{R}), {\bar m}, \bar{s}, \mu_{m},\sigma_{m})$.

There are two common pitfalls with this method. The first is that it assumes that all parameters vary linearly, which is very often not the case, but slight deviations from linearity will only cause minor problems. The second is that the range of each measure will affect the weighting it is given. This second problem can be solved by using the correlation matrix instead of the covariance matrix, which normalises each measure by it's standard deviation. However, after analysing the effectiveness of this algorithm when using the correlation and covariance matrices, we find that these issues cancel each other out. Most of the measures can be reasonably approximated as linear except for $\mu_{m}$ and $\sigma_{m}$ but, when the covariance matrix is used, the weight of these is suppressed because of their much smaller ranges resulting in a better separation of \textbf{Y}-states. We therefore use the covariance matrix in calculating the Principle Components.

\begin{figure*}
\centering
\includegraphics[width=.8\linewidth]{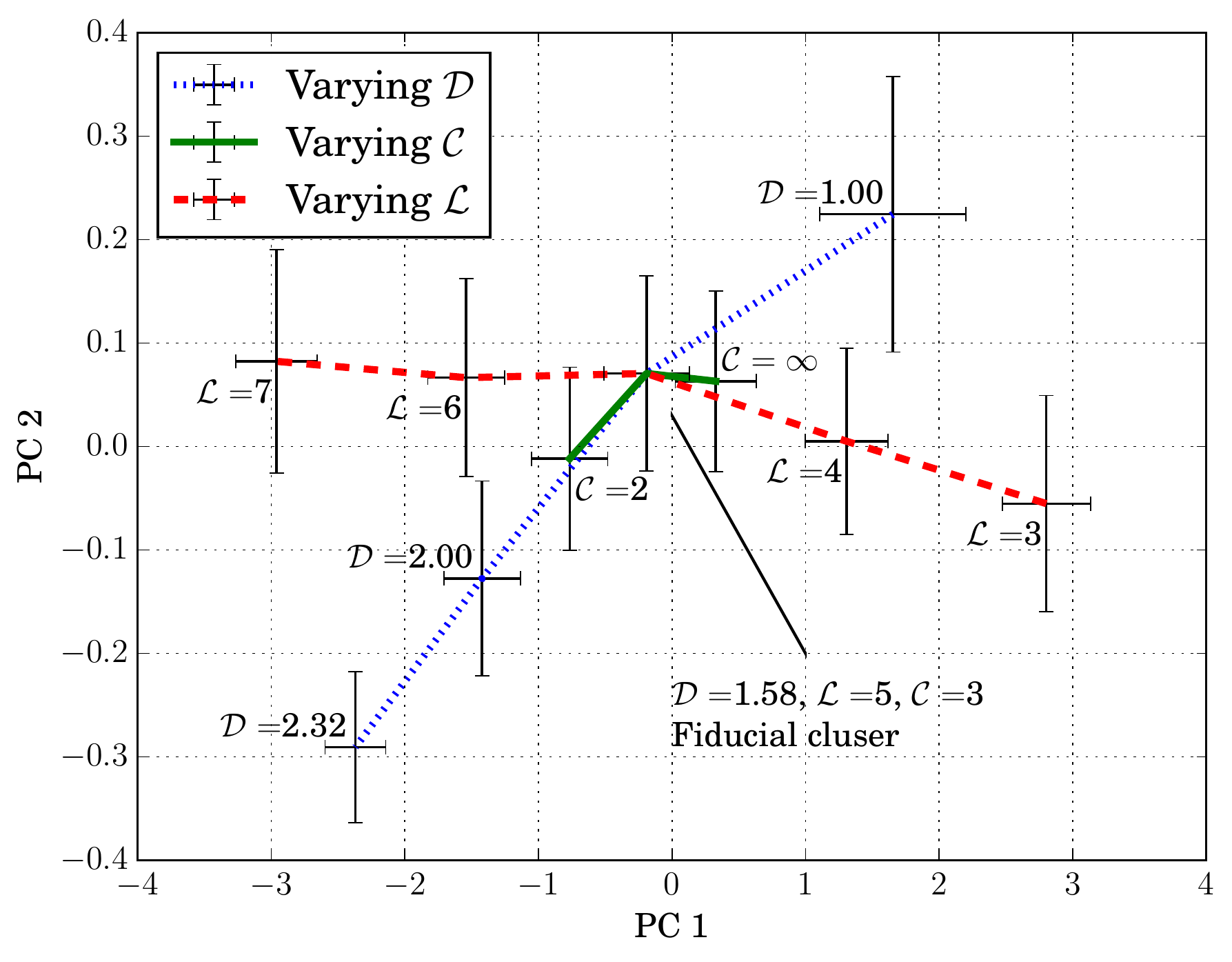}
\caption{The variation of the principle components with the parameters defining the statistical state, (${\cal D}$, ${\cal L}$, ${\cal C}$). The line gives the mean and the error bar gives one standard deviation. Whichever statistical-state parameter is being varied, the other two are held constant at their fiducial values, viz. ${\cal D}=1.58$, ${\cal L}=5$, and ${\cal C}=3$. \change{$\mathcal{D}$ and $\mathcal{L}$ vary almost orthogonally in the PC space, while the variation in $\mathcal{C}$ overlaps with $\mathcal{D}$ and is therefore more difficult to separate.}}
\label{FIG:PCVAR}
\end{figure*}

\begin{figure*}
\centering
\includegraphics[width=2.1\columnwidth]{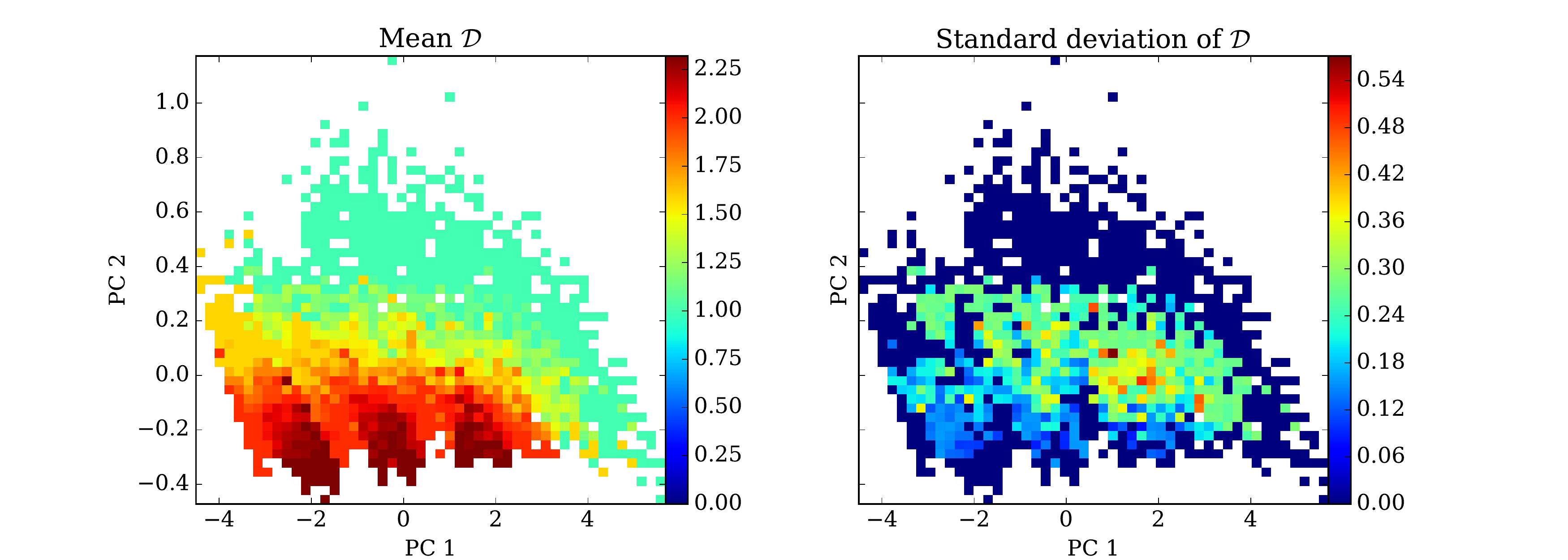}
\includegraphics[width=2.1\columnwidth]{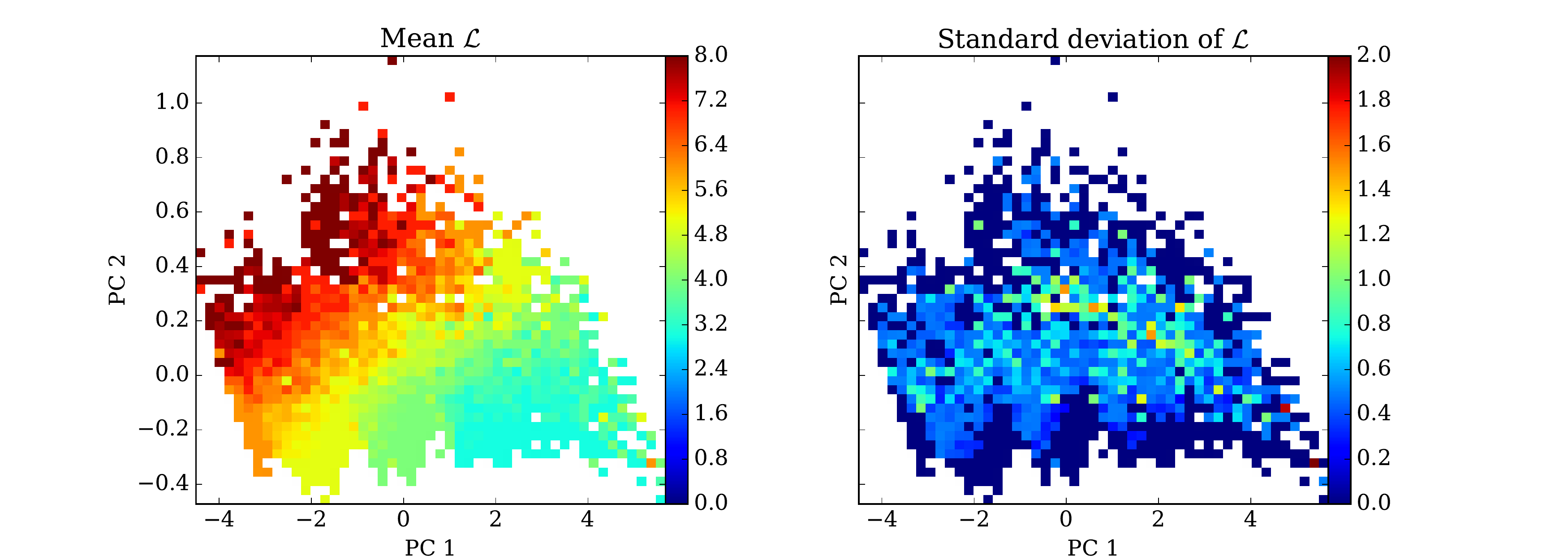}
\caption{The variation of the principle components with ${\cal D}$ (top row) and ${\cal L}$ (bottom row). The plots show the mean value (left hand plots) and standard deviation (right hand plots) of each parameter for each square in a 50x50 grid covering the full parameter space. Areas where there are no data points are white.}
\label{FIG:PCGRID}
\end{figure*}

Each of the 6500 clusters (100 for each of the 65 \textbf{Y} states) is transformed into PC space. 
In order to estimate the parameters of a test (real or artificial) cluster and quantify the uncertainty on this measurement, we grid all the 6500 clusters in PC space on a 50 x 50 regular grid covering the full extent of the data. In each grid square we find all clusters falling in that area, and calculate the mean and standard deviation of their $\mathcal{D}$ and $\mathcal{L}$ values (see figure \ref{FIG:PCGRID}).

The fractal dimension increases somewhat with decreasing PC 1, and increases strongly with decreasing PC 2 (see figure \ref{FIG:PCGRID}, top row). Areas of low ${\cal D}$ have very low $\sigma_{\cal D}$ (in many cases zero) because in this area a small change in parameters makes a large difference in detectable structure, so the different \textbf{Y} values are quite well separated. Middling fractal dimensions (${\cal D} \approxeq 1.58$) have higher errors ($\sigma_{\cal D}>0.3$) because the \textbf{Y} states are less well separated; however even the maximum standard deviation from the mean would only encompass the immediate neighbours in \textbf{Y}-space (${\cal D} = 1.00$ or 2.00). The areas of highest ${\cal D}$ have again lower errors in ${\cal D}$; even though clusters in this area are not well separated, they all have high ${\cal D}$ as it is the edge of the parameter space explored. It should be noted that clusters with  very low ${\cal C}$ also fall in this region, regardless of their $\mathcal{D}$ values, since the larger structures swamp the later generations and erase the signs of substructure.

The number of levels increases with decreasing PC 1 and increasing PC 2 (see figure \ref{FIG:PCGRID}, bottom row). The errors on ${\cal L}$ are low in most areas, particularly on the edges of parameter space where they drop to zero. The high values of $\sigma_{\cal L}$ occurring for PC 1 $>$ 4 and PC 2 $\approx$ 0.2 are caused by small number statistics at the very edge of the parameter space, where test clusters are very unlikely to fall.

${\cal D}$ and ${\cal L}$ both vary systematically across the range of the PCs (see figure \ref{FIG:PCVAR}). \change{${\cal C}$, on the other hand, varies only locally around a particular \textbf{Y}-state, and in the same direction as the variation in $\mathcal{D}$}. Another method is therefore needed to estimate $\mathcal{C}$ once ${\cal D}$ and ${\cal L}$ have been estimated from the principle components.

\newpage

\subsection{Estimating $\mathcal{C}$}\label{SUBSEC:BAYES}%

\begin{figure}
\centering
\includegraphics[width=\columnwidth]{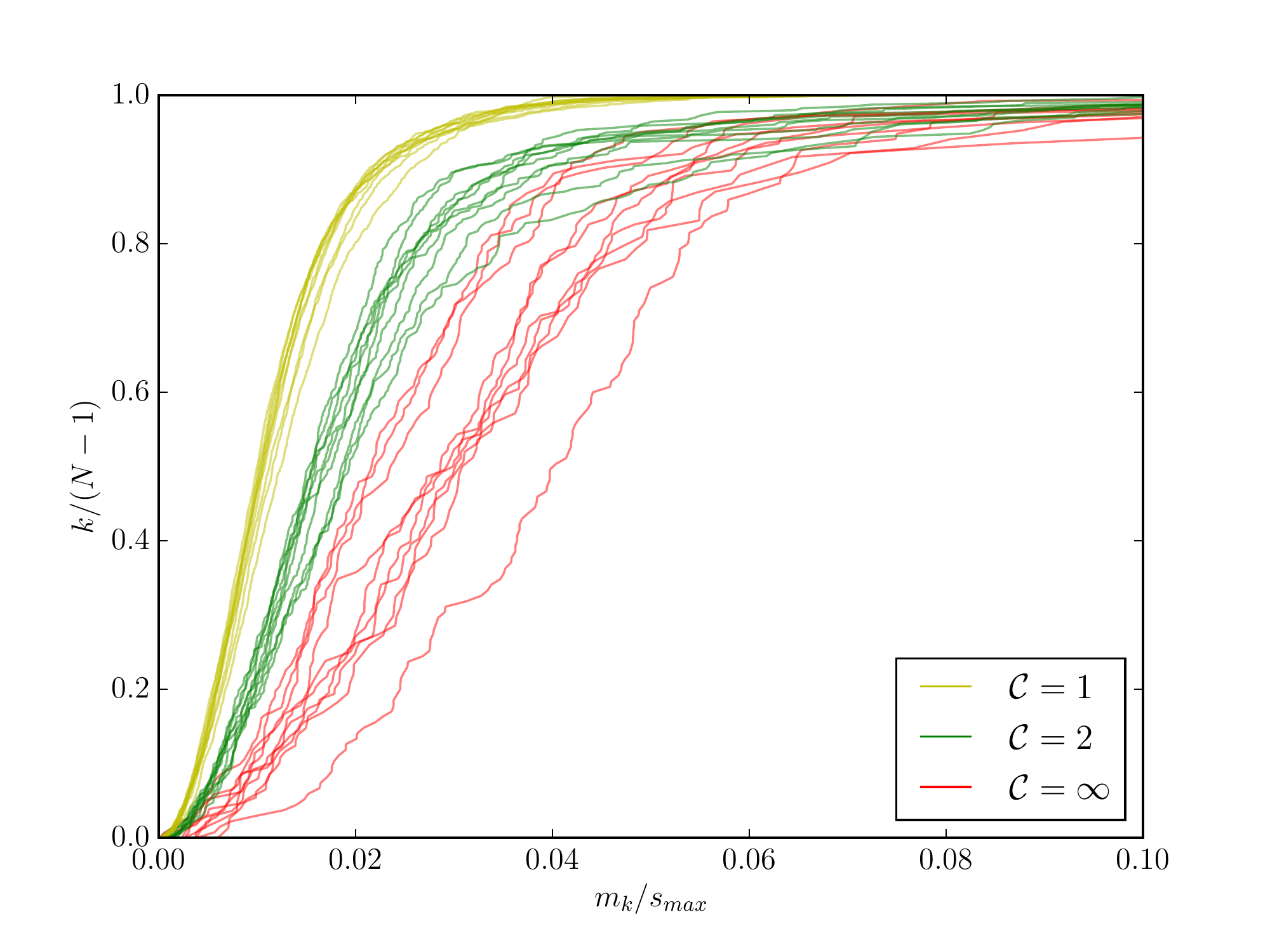}
\caption{The cumulative distribution of minimum spanning tree edges for varying density scaling exponents.}
\label{FIG:AREAMEAS}
\end{figure}

Figure \ref{FIG:AREAMEAS} shows how the cumulative distribution of MST edges (normalised by the size of the cluster, $s_{max}$, and the total number of edges, $\mathcal{N}_{\star}-1$) varies with ${\cal C}$ for set values of $\mathcal{D}$ and $\mathcal{L}$. A cluster with higher $\mathcal{C}$ will have a greater proportion of short edges, shifting this curve to the right. The area above this curve therefore increases with increasing $\mathcal{C}$, but this also varies with $\mathcal{D}$ and $\mathcal{L}$. 

\change{
We use Bayes' theorem to infer the posterior probability of $\mathcal{C}$, given $A$.
\begin{equation}
\label{EQ:BAYES}
P(\mathcal{C}|A) = \frac{P(A|\mathcal{C}) P(\mathcal{C}) }{P(A)}
\end{equation}
The likelihood of a particular $A$, given $\mathcal{C}$ ($P(A|\mathcal{C})$) is calculated from the mean and standard deviation of $A$ ($\mu_{A_{\mathcal{C}}}, \sigma_{A_{\mathcal{C}}}$) over the 100 realisations at each \textbf{Y}-state,
\begin{equation}
P(A|\mathcal{C}) = \frac{1}{\sigma\sqrt{2\pi}}  e^{-0.5 (\frac{A-\mu_{A_{\mathcal{C}}}}{\sigma_{A_{\mathcal{C}}}})}
\end{equation}
$P(\mathcal{C})$ is a weight given to each value of $\mathcal{C}$ based on prior knowledge of the distribution. In this case, each value of $\mathcal{C}$ is given equal weight. $P(A)$ is a normalisation constant to ensure that the probabilities add up to unity.
We calculate the expected value of $\mathcal{C}$ and its standard deviation from the posterior probabilities, i.e.
\begin{eqnarray}
E_{\mathcal{C}} & = & \frac{\sum P(\mathcal{C}|A) \mathcal{C}}{\sum P(\mathcal{C}|A)}\\
\sigma_{\mathcal{C}} & = & \frac{\sum P(\mathcal{C}|A) (\mathcal{C}-E_{\mathcal{C}})^{2}}{\sum P(\mathcal{C}|A)}
\end{eqnarray}
}

\section{Evaluation and application}\label{SEC:APPLY}%

Given the 2D projection of a real, synthetic or simulated cluster, we first compute the six measures, ${\bf Z}=({\log(\mathcal{N}_\star)},{\log(\mathcal{R})},{\bar s},{\bar m},{\mu_{m}},{\sigma_{m}})$ and these are transformed into principle components. The values of the PCs identify which grid square the test cluster falls into \change{(see figure \ref{FIG:PCGRID})} and the mean and standard deviation of artificial clusters in that grid square give an estimate and uncertainty for the $\mathcal{D}$ and $\mathcal{L}$ values of the test cluster. We then compare the value of $A$ for the test cluster to the analytic clusters with the same $\mathcal{D}$ and $\mathcal{A}$ and use a Bayesian approach to estimate $\mathcal{C}$ and it's error.

\subsection{Evaluation with synthetic star clusters}\label{SUBSEC:SYNTH}%

We focus our attention on clusters with ${\cal D}=1.00,\;1.58\;{\rm and}\;2.00$, ${\cal L}=4,\;5\;{\rm and}\;6$ and ${\cal C}=2,\;3\;{\rm and}\;\infty$. For each of these 27 \textbf{Y}-states, we have created 10 synthetic fractal star clusters, projected them at a random angle, and analysed them using the algorithm described in Section \ref{SEC:METHOD}. We find the means and standard deviations of ${\cal D}_{_{\rm OUT}}/{\cal D}_{_{\rm IN}}$,  ${\cal L}_{_{\rm OUT}}/{\cal L}_{_{\rm IN}}$,  ${\cal C}_{_{\rm OUT}}/{\cal C}_{_{\rm IN}}$, where $Y_{_{i,\,{\rm IN}}}$ is the value of $Y_i$ that went into the construction of a synthetic three-dimensional star cluster, and $Y_{_{i,\,{\rm OUT}}}$ is the value of $Y_i$ estimated from the projected two-dimensional image of this cluster. These values are given in Table \ref{TAB:InOut}.

We can see that $\mathcal{D}$ and $\mathcal{L}$ are reliably estimated across the parameter space, although a low $\mathcal{D}$ is often overestimated when $\mathcal{L}$ is low. $\mathcal{C}$ is not as well constrained, particularly when $\mathcal{C}=\infty$. This could be improved by creating more analytic clusters with different values of $\mathcal{C}$ to give more \textit{a priori} information for the Bayesian analysis, but as \change{this gets} fairly computationally intensive we leave this to future work. We hope that use of this algorithm will reveal the areas of parameter space populated by real and simulated clusters, and therefore we can focus our attention for improvements in these regions.

\begin{table*}
\centering
\tabcolsep=0.18cm
\begin{tabular}{r|ccc|ccc|ccc}\hline

&&&&&&&&& \\

${\cal D}_{_{\rm IN}}=$ & &  & & & $1.00$ & & &  & \\
${\cal L}_{_{\rm IN}}=$ & & 4 & & & 5 & & & 6 & \\
${\cal C}_{_{\rm IN}}=$ & 2 & 3 & $\infty$ & 2 & 3 & $\infty$ & 2 & 3 & $\infty$ \\

${\cal D}_{_{\rm OUT}}/{\cal D}_{_{\rm IN}}=$ & 
$1.37\!\pm\!0.19$ & $1.18\!\pm\!0.15$ & $1.00\!\pm\!0.01$ & $1.19\!\pm\!0.13$ & $1.11\!\pm\!0.09$ & $1.07\!\pm\!0.09$ & $1.25\!\pm\!0.13$ & $1.11\!\pm\!0.19$ & $1.23\!\pm\!0.23$  \\ 
${\cal C}_{_{\rm OUT}}/{\cal C}_{_{\rm IN}}=$ & 
$0.47\!\pm\!3.52$ & $0.99\!\pm\!2.26$ & ($2.53\!\pm\!0.55$) & $0.56\!\pm\!3.03$ & $1.05\!\pm\!2.16$ & ($5.82\!\pm\!2.30$) & $0.66\!\pm\!2.48$ & $0.80\!\pm\!3.29$ & ($7.06\!\pm\!2.47$)   \\ 
${\cal L}_{_{\rm OUT}}/{\cal L}_{_{\rm IN}}=$ & 
$0.99\!\pm\!0.10$ & $0.90\!\pm\!0.09$ & $0.90\!\pm\!0.07$ & $1.05\!\pm\!0.10$ & $0.98\!\pm\!0.11$ & $0.96\!\pm\!0.15$ & $0.96\!\pm\!0.08$ & $0.98\!\pm\!0.15$ & $0.79\!\pm\!0.11$   \\

&&&&&&&&& \\

${\cal D}_{_{\rm IN}}=$ & &  & & & $1.58$ & & &  & \\
${\cal L}_{_{\rm IN}}=$ & & 4 & & & 5 & & & 6 & \\
${\cal C}_{_{\rm IN}}=$ & 2 & 3 & $\infty$ & 2 & 3 & $\infty$ & 2 & 3 & $\infty$ \\

${\cal D}_{_{\rm OUT}}/{\cal D}_{_{\rm IN}}=$ & 
$1.07\!\pm\!0.15$ & $1.08\!\pm\!0.17$ & $0.89\!\pm\!0.15$ & $1.00\!\pm\!0.15$ & $0.96\!\pm\!0.11$ & $0.93\!\pm\!0.15$ & $1.01\!\pm\!0.09$ & $1.00\!\pm\!0.11$ & $0.98\!\pm\!0.11$  \\ 
${\cal C}_{_{\rm OUT}}/{\cal C}_{_{\rm IN}}=$ & 
$0.29\!\pm\!3.63$ & $1.02\!\pm\!2.34$ & ($6.20\!\pm\!2.45$) & $0.68\!\pm\!3.33$ & $1.11\!\pm\!2.39$ & ($4.93\!\pm\!3.44$) & $0.63\!\pm\!3.18$ & $1.11\!\pm\!2.23$ & ($4.25\!\pm\!3.15$) \\ 
${\cal L}_{_{\rm OUT}}/{\cal L}_{_{\rm IN}}=$ & 
$1.07\!\pm\!0.12$ & $0.92\!\pm\!0.13$ & $1.03\!\pm\!0.15$ & $1.06\!\pm\!0.13$ & $1.02\!\pm\!0.10$ & $0.99\!\pm\!0.13$ & $1.07\!\pm\!0.07$ & $0.99\!\pm\!0.06$ & $0.94\!\pm\!0.09$ \\

&&&&&&&&& \\

${\cal D}_{_{\rm IN}}=$ & &  & & & $2.00$ & & &  & \\
${\cal L}_{_{\rm IN}}=$ & & 4 & & & 5 & & & 6 & \\
${\cal C}_{_{\rm IN}}=$ & 2 & 3 & $\infty$ & 2 & 3 & $\infty$ & 2 & 3 & $\infty$ \\

${\cal D}_{_{\rm OUT}}/{\cal D}_{_{\rm IN}}=$ & 
$0.94\!\pm\!0.08$ & $0.93\!\pm\!0.10$ & $0.91\!\pm\!0.12$ & $0.96\!\pm\!0.08$ & $0.92\!\pm\!0.09$ & $1.00\!\pm\!0.10$ & $0.96\!\pm\!0.06$ & $1.03\!\pm\!0.08$ & $1.06\!\pm\!0.08$ \\ 
${\cal C}_{_{\rm OUT}}/{\cal C}_{_{\rm IN}}=$ & 
$0.83\!\pm\!2.48$ & $0.83\!\pm\!2.60$ & ($3.30\!\pm\!3.54$) & $0.83\!\pm\!1.75$ & $0.72\!\pm\!3.53$ & ($7.48\!\pm\!3.70$) & $0.58\!\pm\!2.23$ & $0.85\!\pm\!2.41$ & ($6.05\!\pm\!3.61$)  \\ 
${\cal L}_{_{\rm OUT}}/{\cal L}_{_{\rm IN}}=$ & 
$1.10\!\pm\!0.08$ & $1.07\!\pm\!0.09$ & $1.05\!\pm\!0.09$ & $1.06\!\pm\!0.10$ & $1.07\!\pm\!0.08$ & $0.97\!\pm\!0.09$ & $1.04\!\pm\!0.08$ & $0.97\!\pm\!0.08$ & $0.93\!\pm\!0.08$ \\

&&&&&&&&& \\\hline

\end{tabular}
\caption{Means and standard deviations for the ratios between (i) the statistical-state parameters used in the construction of synthetic 3D fractal star clusters, $Y_{_{i,\,{\rm IN}}}$, and (ii) the statistical-state parameters, $Y_{_{i,\,{\rm OUT}}}$, derived from 2D projections of these clusters. \change{In cases where $\mathcal{C}=\infty$ this ratio is meaningless, but we use a very high number in place of infinity so these results are presented for completeness.}}
\label{TAB:InOut}
\end{table*}

\subsection{Application to observed star clusters}\label{SUBSEC:OBS}%

\begin{figure}
\centering
\includegraphics[width=\columnwidth]{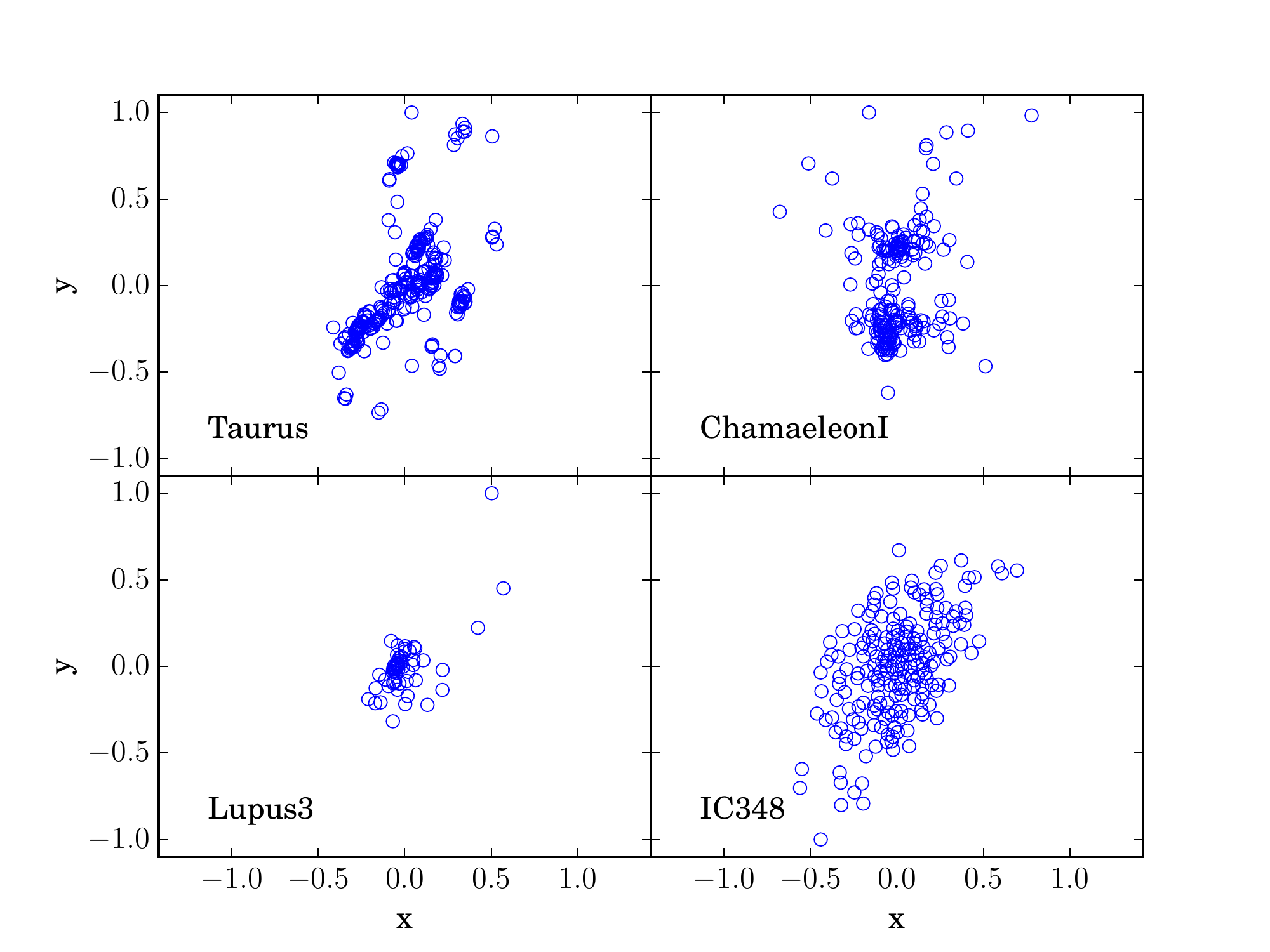}
\caption{Young stellar objects in Taurus, Chamaeleon I, Lupus 3 and IC 348. Each cluster has been normalised so that the origin is the mean position of all the stars and the radius of the cluster (distance from the mean position to the furthest star) is unity.}
\label{FIG:REAL}
\end{figure}

We have tested this algorithm on 4 clusters of young stellar objects taken from \cite{2011K+M}; Lupus~3, Taurus, Chamaeleon~I and IC~348. \change{These are shown in figure \ref{FIG:REAL}}

\subsubsection{Binaries in real clusters}\label{SUBSUBSEC:BINARIES}%

\begin{figure}
\centering
\includegraphics[width=\columnwidth]{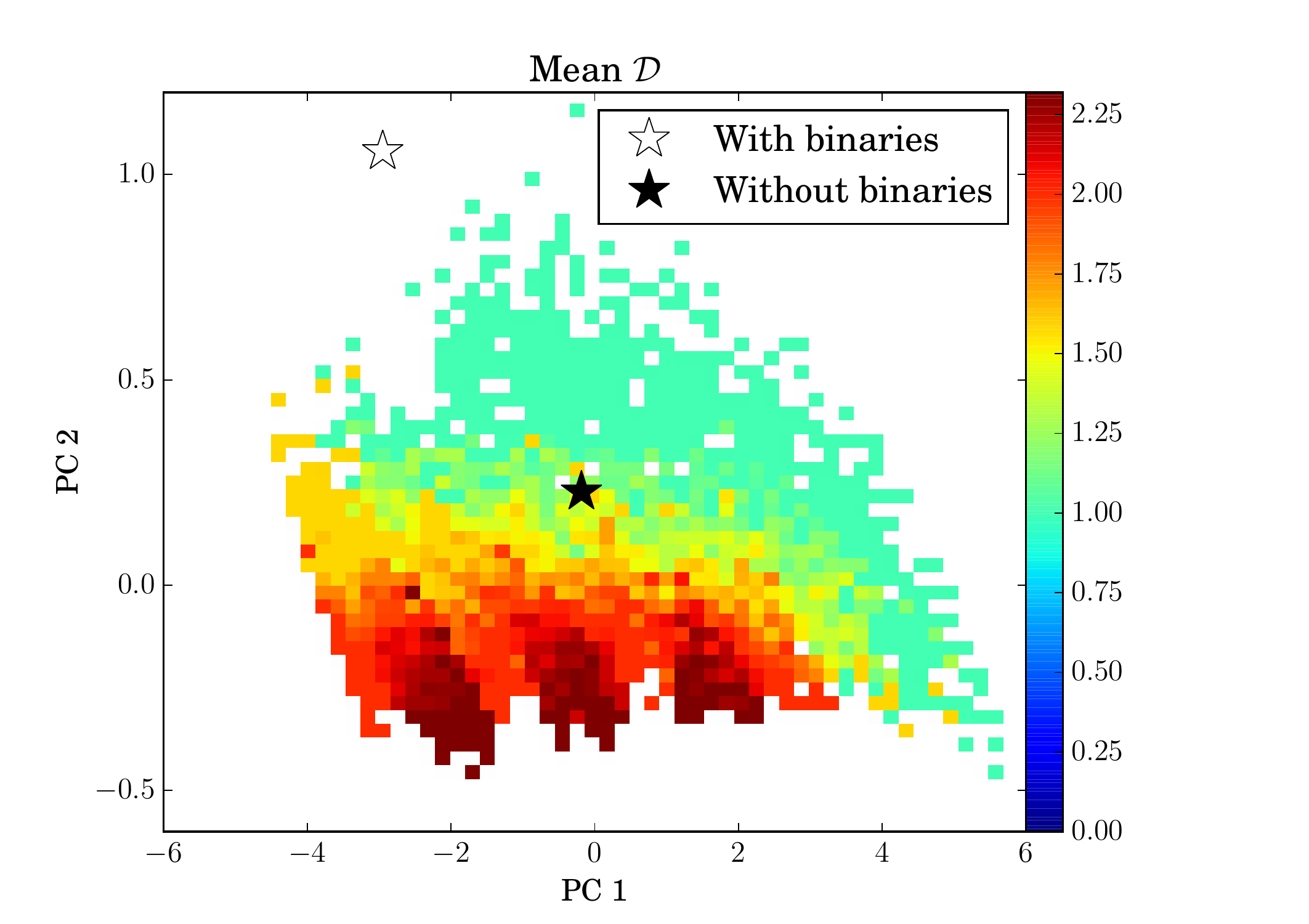}

\caption{The Chamaeleon I cluster with (open star) and without (filled star) binaries superimposed on the $\mathcal{L}$ grid. It lies outside the parameter space if binaries are left in, but moves into the area of fractal clusters when binaries are removed.}
\label{FIG:REALBIN}
\end{figure}

\change{The artificial clusters generated for this analysis model various types of hierarchical clustering. However, one major difference in the structure of real star clusters that is not modelled in this work is binary or higher order multiple systems). If these are present in a cluster, they will produce many of the shortest edges in both the complete graph and the minimum spanning tree, and will therefore significantly skew some of the measures. In the 4 real clusters analysed here the effect of removing binaries decreases $\log(\mathcal{R})$ by $\approx$ 30\% and increases $\mu_{m}$ by $\approx$ 20\%.}

Figure \ref{FIG:REALBIN} shows the effect of binary systems in a real cluster on it's position in Principle Component space. Using the original cluster data from \cite{2011K+M}, most of the real clusters lie well outside the parameter space of artificial fractal clusters. Once binaries have been removed, all real clusters lie close to the parameter space explored and show a range of properties from Lupus 3 and Taurus (low fractal dimension and high number of levels) to Chamaeleon I (lower number of levels but still low $\mathcal{D}$) to IC 348 which appears at the high-$\mathcal{D}$ limit of the parameter space, indicating a smoother distribution rather than a fractal sub-clustering.

\cite{1995L} discussed the relation between hierarchical clustering and multiple systems and found that they showed a distinctly different distribution of separations. \cite{1999GKBW} examined this in several clusters and found a characteristic separation of $\approx 0.03$ pc that distinguishes hierarchical clustering from the regime of binary and multiple systems. We use this to remove the effect of binaries from real clusters before analysis. 

Pairs of stars separated by less than 0.03 pc are classified as binaries. However, some pairs of stars with small enough separations to be defined as \textquotedblleft binary" will be generated when an artificial cluster is viewed in two dimensions, due to chance alignments of stars which are well separated in the third dimension. Numerical experiments with artificial clusters across the parameter space show that this is simply dependent on average surface density:
\begin{equation}
\log (N_{bin}) = 1.86 \log(\mathcal{N}_{\star}/\pi R_{cluster}^{2}) - 0.56
\end{equation}
where $N_{bin}$ is the number of chance alignments smaller than 0.03 pc and $R_{cluster}$ is the maximum distance of any star from the mean position of all the stars. This can be easily calculated for a real cluster. When small separations are found in a real cluster they are randomly removed until the number does not exceed the predicted $N_{bin}$, thereby removing the effect of binaries on the statistical measures but without removing the proportion of small separations expected from projection. In the case of higher order multiples, this will result in the whole system being replaced by a single star. In effect, we are studying the hierarchical distribution of systems, rather than of stars.

\subsubsection{Results of analysis of real clusters}\label{SUBSUBSEC:REAL}

\begin{figure}
\centering
\includegraphics[width=\columnwidth]{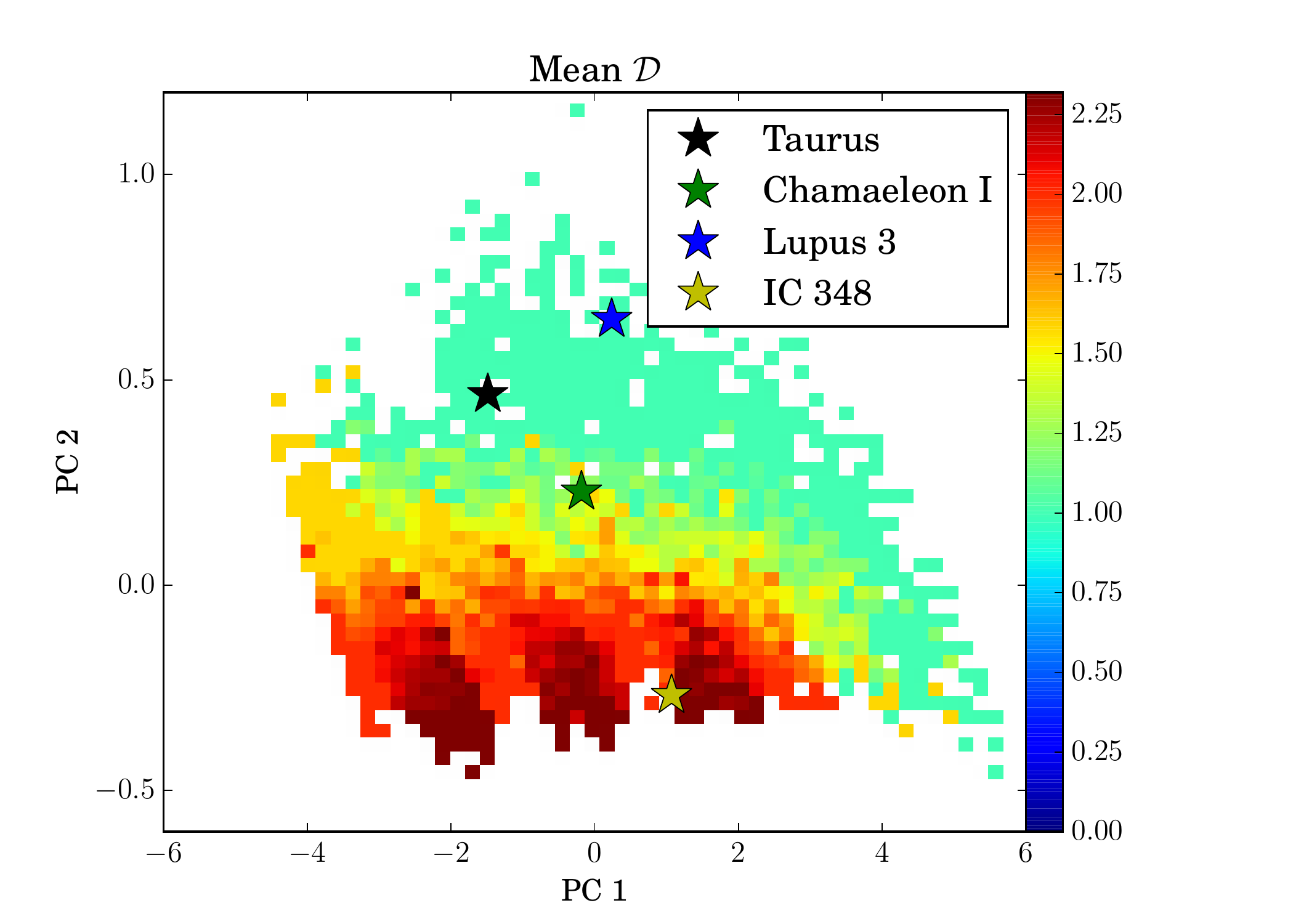}
\caption{Real clusters superimposed on the background of mean ${\cal D}$.}
\label{FIG:REALPC}
\end{figure}

Figure \ref{FIG:REALPC} shows the placement of the 4 real clusters in relation to the artificial cluster parameter space after binaries have been removed. The estimated parameters of the 4 clusters are given in table \ref{TAB:RealResults} along with previous estimates of their structure. 

\change{The estimated parameters appear to be compatible with a visual inspection of the data (see figure \ref{FIG:REAL}) whilst having the merit of being objective and quantitative, whereas visual inspection is subjective and qualitative. Both Taurus and Chamaeleon~I  show clear sub-structure, but Taurus appears to be more clumpy, so has lower $\mathcal{D}$ than Chamaeleon~I. Taurus is the largest of these 4 regions, so the sub-structure has a greater range of scales. This is reflected in the estimates of $\mathcal{L}$. The estimated fractal dimensions are lower than those obtained using the original $\mathcal{Q}$ parameter, but the trend is the same. IC~348 was classified by CW04 as not having sub-structure but being centrally concentrated. Our analysis agrees with this, although we are not able to quantify the structure, only to say that it is not measurably sub-structured. Lupus~3 was not analysed in CW04, but our analysis shows it to be highly sub-structured, but over a smaller range of scales that Taurus (lower $\mathcal{L}$). It also has a higher $\mathcal{C}$ (or lower $\mathcal{C}^{-1}$), which reflects the fact that there are very few stars outside the main dense clump, while Taurus and Chamaeleon~I have more evenly distributed stars outside the densest regions.}

\begin{table}
\begin{center}
\begin{tabular}{rcccc}\hline
&&&& \\
Cluster:		& Lupus 3 		& Chamaeleon I 		& Taurus 			& IC 348 \\
${\cal D}=$	 	& $1.0 \pm 0.0$ & $1.5 \pm 0.2$ 	& $1.0 \pm 0.0$ 	& \sc{Not Frac.} \\
${\cal L}=$ 	& $7.0 \pm 0.3$ & $5.2 \pm 0.4$	 	& $8.0 \pm 0.0$ 	& -- \\
${\cal C}^{-1}=$& $0.1 \pm 0.0$ & $0.4 \pm 0.0$ 	& $0.4 \pm 0.0$ 	& -- \\
CW04 			& N/A 			& ${\cal D}$ = 2.25 & ${\cal D}$ = 1.5 	&  $\alpha$ = 2.2 \\
&&&& \\\hline
\end{tabular}
\caption{Parameters estimated for the four real clusters after the removal of binary or multiple systems. The fourth row gives the results quoted in CW04 using the original $\mathcal{Q}$ parameter method, where $\mathcal{D}$ is the fractal dimension and $\alpha$ is the radial density exponent.}
\label{TAB:RealResults}
\end{center}
\end{table}

\section{Conclusions}\label{SEC:CONC}%

We present a new algorithm for quantifying the fractal nature of young star clusters in terms of three parameters:
\begin{itemize}
\item the fractal dimension, $\mathcal{D}$ - a measure of the clumpiness or smoothness of the distribution;
\item the number of levels, $\mathcal{L}$ - a measure of the range of scales of substructure within the overall cloud;
\item the density scaling exponent, $\mathcal{C}$ - a measure of the relative distribution of mass on the different scales.
\end{itemize}
It is able to reliably classify the internal structure of young stellar clusters within a parameter space bounded by the following limitations:
\begin{enumerate}
\item $\mathcal{L} \leq$ 3 will not have enough substructure to be detectably fractal.
\item $\mathcal{D} \geq$ 2.32 will fill most of the area when projected into 2D, and therefore appear smooth.
\item $\mathcal{C} \leq$ 1 will overpopulate the higher levels and swamp any substructure on smaller scales.
\end{enumerate}

The estimated properties of Taurus, Lupus~3, Chamaeleon~I and IC~348 fit with a visual assessment of their structure, and the new method reduces problems encountered using the old $\mathcal{Q}$ parameter due to not considering all 3 parameters inherent in an artificially generated fractal cluster.

We anticipate that this method will be useful for:
\begin{itemize}
\item quantitative analysis of large numbers of structures in the huge data sets available from modern observing methods, to avoid the necessity for visual inspection;
\item unbiased analysis of the results of simulations in 2 or 3 dimensions;
\item analytical comparison of observational and simulated data sets to validate results and inform inferences about the similarity of observed regions to simulated environments.
\end{itemize}

\change{The algorithm described in this paper will shortly be available at \url{https://github.com/SJaffa/Q_plus}.}

\section*{Acknowledgements}%

SEJ gratefully acknowledges the support of a postgraduate scholarship from the School of Physics \& Astronomy at Cardiff University. APW and OL gratefully acknowledge the support of the consolidated grant ST/K00926/1 from the UK Science and Technology Facilities Council, and of the EU-funded {\sc vialactea} Network {\sc fp}7-{\sc space}-607380. The computations were performed on the Cardiff University Advanced Research Computing facility, {\sc arcca}.

\bibliographystyle{mn2e}
\bibliography{antsrefs,papers}

\vspace{1.0cm}

\label{lastpage}
\end{document}